\documentclass[12pt,epsfig]{article}
\usepackage{graphicx}
\usepackage{amssymb}
\usepackage{mathrsfs}

\newcommand{\beq}[1] {\begin{equation}\label{#1} }
\newcommand{\eeq} {\end{equation} }
\newcommand{\bea}[1]{\begin{eqnarray}\label{#1} }
\newcommand{\eea}{\end{eqnarray}}

\begin{document}

\vspace*{-0.5cm}
\begin{flushright}
OSU-HEP-06-6
\end{flushright}
\vspace{0.5cm}

\begin{center}
{\Large {\bf A New Two Higgs Doublet Model} }

\vspace*{1.5cm} S. Gabriel\footnote{Email address: svengab@msn.com}
and S. Nandi\footnote{Email address: s.nandi@okstate.edu}

\vspace*{0.5cm}
{\it Department of Physics and Oklahoma Center for High Energy Physics,\\
Oklahoma State University, Stillwater, Oklahoma, 74078\\}

\end{center}

\begin{abstract}

We propose an extension of the Standard Model by extending the EW
symmetry to $SU(2)_L \times U(1) \times Z_2$ and introducing three
$SU(2) \times U(1)$ singlet right handed neutrinos, $N_R$, and an
additional Higgs doublet, $\phi$.  While the SM gauge bosons and the
quarks and charged leptons acquire masses from the spontaneous
breaking of $SU(2)_L \times U(1)$ symmetry at the electroweak scale,
the neutrinos acquire masses from the spontaneous breaking of the
discrete $Z_2$ symmetry at a scale of $10^{-2}$ eV.  In addition to
providing a new mechanism for generating tiny masses for the
neutrinos, the model has interesting implications for neutrinoless
double beta decay and the Higgs signals at high energy colliders, as
well as in astrophysics and cosmology.

\end{abstract}

\section{Introduction}

All experimental results to date are in excellent agreement with the
predictions of the Standard Model (SM).  However, one essential
ingredient of the SM, the existence of the Higgs boson, as well as
the interactions of the Higgs boson with the SM particles, are yet
to be experimentally established.  In the SM, we have only one Higgs
doublet, and its interactions with the gauge bosons and fermions are
completely determined by theory.  However, there are several
extensions of the SM  Higgs sector which include more than one
electroweak (EW) doublet, as well as EW singlets \cite{Higgs
Hunters}.  Such extensions include the Minimal Supersymmetric
Standard Model (MSSM) which requires two EW Higgs doublets, and the
Next to Minimal Supersymmetric Standard Model (NMSSM) which has two
doublets and one singlet, as well as models with more than one
singlet \cite {singlets}.  There are also non-supersymmetric two
Higgs doublet models. Typically, all these extensions involve new
additional parameters, and as a result, are not as predictive as the
SM for the interactions of the Higgs bosons with the SM gauge bosons
and fermions.

Although the mass of the Higgs boson is not predicted by the SM,
accurate measurements of the top quark and the $W$ boson mass at the
Tevatron, as well as the $Z$ boson mass at LEP, have narrowed the SM
Higgs boson mass between 80 and 200 GeV \cite{EWprecision}.  Failure
to observe the SM Higgs boson at LEP2 has also placed a direct lower
bound of 114 GeV on its mass \cite{higgs_search}.  The dominant
decay modes of the SM Highs boson are to $b\overline{b}$, $WW$, $ZZ$
or $t\overline{t}$, depending on its mass.  The extensions of the SM
may avoid constraints on the Higgs mass, and may allow Higgs bosons
with masses less than the above limits.  The dominant decay modes of
the Higgs bosons can also be altered in such extensions, thus
transforming the discovery signals for the Higgs bosons at the Large
Hadron Collider (LHC).

Another fundamental question is how the SM neutrinos acquire tiny
masses that are almost a billionth times smaller than the quark and
charged lepton masses and whether the neutrinos are Majorana or
Dirac particle.  The most popular explanation of the tiny neutrino
masses is the sea-saw mechanism \cite{see saw} which invokes the
existence of a very heavy EW singlet ($\sim 10^{14}$ GeV)
right-handed neutrino.  Although there are several indirect benefits
of its existence, there is no direct experimental evidence for such
a heavy particle.  It is important to explore other possibilities to
explain the tiny neutrino masses \cite{neu mass}.

In this work, we present an alternate explanation for the tiny
masses of the SM neutrinos, as well as possibilities for altering
signals for discovery of the Higgs at the LHC.  Our proposal is to
extend the SM electroweak symmetry to  $SU(2)_L \times U(1) \times
Z_2$ and introduce three $SU(2) \times U(1)$ singlet right handed
neutrinos, $N_R$, as well as an additional Higgs doublet, $\phi$.
While the SM symmetry is spontaneously broken by the VEV of an EW
doublet $\chi$ at the 100 GeV scale, the discrete symmetry $Z_2$ is
spontaneously broken by the tiny VEV of this additional doublet
$\phi$ at a scale of $10^{-2}$ GeV. Thus in our model, tiny neutrino
masses are related to this $Z_2$ breaking scale. We note that
although our model has extreme fine tuning, that is no worse than
the fine tuning problem in the usual GUT model. Many versions of the
two Higgs doublet model have been extensively studied in the past
\cite{Higgs Hunters}. The examples include:  a) a supersymmetric two
Higgs doublet model, b) non-supersymmetric two Higgs doublet models
i) in which both Higgs doublets have vacuum expectation values
(VEV's) with one doublet coupling to the up type quarks only, while
the other coupling to the down type quarks only, ii) only one
doublet coupling to the fermions, and iii) only one doublet having
VEV's and coupling to the fermions \cite{inert}. What is new in our
model is that one doublet couples to all the SM fermions except the
neutrinos, and has a VEV which is same as the SM VEV, while the
other Higgs doublet couples only to the neutrinos with a tiny VEV
$\sim 10^{-2}$ eV. This latter involves the Yukawa coupling of the
left-handed SM neutrinos with a singlet right-handed neutrino,
$N_R$.  The left-handed SM neutrinos combine with the singlet
right-handed neutrinos to make massive Dirac neutrinos. The neutrino
mass is so tiny because of the tiny VEV of the second Higgs doublet,
which is responsible for the spontaneous breaking of the discrete
symmetry, $Z_2$.  Note that in the neutrino sector, our model is
very distinct from the sea-saw model. Lepton number is strictly
conserved, and hence no $N_R N_R$ mass terms are allowed. Thus the
neutrino is a Dirac particle, and there is no neutrinoless double
$\beta$ decay in our model.  In the Higgs sector, in addition to the
usual massive neutral scalar and pseudoscalar Higgs, and two charged
Higgs, our model contains one essentially massless scalar Higgs.  We
will show that this is still allowed by the current experimental
data and can lead to an invisible decay mode of the SM-like Higgs
boson, thus complicating the Higgs searches at the Tevatron and the
LHC.

\section{Model and the Formalism}
Our proposed model is based on the symmetry group $SU(3)_c \times
SU(2)_L \times U(1) \times Z_2$.  In addition to the usual SM
fermions, we have three EW singlet right-handed neutrinos, $N_{Ri},
i=1-3$, one for each family of fermions.  The model has two Higgs
doublets, $\chi$ and $\phi$.  All the SM fermions and the Higgs
doublet $\chi$, are even under the discrete symmetry, $Z_2$, while
the RH neutrinos and the Higgs doublet $\phi$ are odd under $Z_2$.
Thus all the SM fermions except the left-handed neutrinos, couple
only to $\chi$. The SM left-handed neutrinos, together with the
right-handed neutrinos, couple only to the Higgs doublet $\phi$. The
gauge symmetry $SU(2) \times U(1)$ is broken spontaneously at the EW
scale by the VEV of $\chi$, while the discrete symmetry $Z_2$ is
broken by a VEV of $\phi$, and we take $\langle\phi\rangle \sim
10^{-2}~eV$. Thus, in our model, the origin of the neutrino masses
is due to the spontaneous breaking of the discrete symmetry $Z_2$.
The neutrinos are  massless in the limit of exact $Z_2$ symmetry.
Through their Yukawa interactions with the Higgs field $\phi$, the
neutrinos acquire masses much smaller than those of the quarks and
charged leptons due to the tiny VEV of $\phi$.

The Yukawa interactions of the Higgs fields with the leptons are
\bea{fermion}
 L_Y =y_{l}\overline{\Psi}^{l}_{L}l_{R}\chi+y_{\nu_{l}}\overline{\Psi}^{l}_{L}N_{R}\widetilde{\phi}+h.c.,
 \eea
where $\overline{\Psi}^{l}_{L} =
(\overline{\nu}_{l},~\overline{l})_L$ is the usual lepton doublet
and $l_R$ is the charged lepton singlet.  The first term gives rise
to the mass of the charged leptons, while the second term gives a
tiny neutrino mass.  The interactions with the quarks are the same
as in the Standard Model with $\chi$ playing the role of the SM
Higgs doublet. Note that in our model, a SM left-handed neutrino,
$\nu_L$ combines with a right handed neutrino, $N_R$, to make a
massive Dirac neutrino with a mass $\sim 10^{-2}$ eV, the scale of
$Z_2$ symmetry breaking.

For simplicity, we do not consider CP violation in the Higgs sector.
(Note that in this model, spontaneous CP violation would be highly
suppressed by the small VEV ratio and could thus be neglected.
However, one could still consider explicit CP violation).  The most
general Higgs potential consistent with the $SM \times Z_2$ symmetry
is \cite{sn}

\bea{potential}
\ V =
-\mu^2_1~\chi^{\dag}\chi-\mu^2_2~\phi^{\dag}\phi+\lambda_1(\chi^{\dag}\chi)^{2}+\lambda_2(\phi^{\dag}\phi)^{2}+\lambda_3(\chi^{\dag}\chi)(\phi^{\dag}\phi)-\lambda_4|\chi^{\dag}\phi|^{2}\nonumber\\-\frac{1}{2}\lambda_5[(\chi^{\dag}\phi)^{2}+(\phi^{\dag}\chi)^{2}].
\eea
The physical Higgs fields are a charged field $H$, two neutral
scalar fields $h$ and $\sigma$, and a neutral pseudoscalar field
$\rho$.  In the unitary gauge, the two doublets can be written

\bea{chi} \chi = \frac{1}{\sqrt{2}}\left(
                                               \begin{array}{c}
                                                 \sqrt{2} (V_\phi/V)H^{+} \\
                                                 h_0 + i (V_\phi/V)\rho
                                                 +V_\chi\\
                                               \end{array}
                                             \right),\nonumber
                                             \eea
\bea{phi} \phi = \frac{1}{\sqrt{2}}\left(
                                               \begin{array}{c}
                                                 -\sqrt{2} (V_\chi/V)H^{+} \\
                                                 \sigma_0 - i (V_\chi/V)\rho
                                                 +V_\phi\\
                                               \end{array}
                                             \right),
 \eea
 where $V_\chi = \langle\chi\rangle$, $V_\phi = \langle\phi\rangle$,
 and $V^{2} = V^{2}_\chi + V^{2}_\phi$.  The particle masses are

 \bea{masses} m^2_{W} = \frac{1}{4}g^{2}V^{2},~ m^2_{H} = \frac{1}{2}(\lambda_4 +\lambda_5)V^{2},~
 m^{2}_\rho = \lambda_5 V^{2},\nonumber
\eea
 \bea{more}m^{2}_{h,\sigma} = (\lambda_1
 V^{2}_\chi +\lambda_2 V^{2}_\phi)\pm \sqrt{(\lambda_1
 V^{2}_\chi -\lambda_2 V^{2}_\phi)^{2}
 +(\lambda_3-\lambda_4-\lambda_5)^{2} V^{2}_\chi V^{2}_\phi}.
 \eea
An immediate consequence of the scenario under consideration is a
very light scalar $\sigma$ with mass

\bea{light} m^{2}_\sigma = 2\lambda_2
V^{2}_\phi[1+O(V_\phi/V_\chi)]. \eea
The mass eigenstates $h,
\sigma$ are related to the weak eigenstates $h_0, \sigma_0$ by

\bea{states}
 h_0 = ch+s\sigma,~\sigma_0 = -sh+c\sigma,
  \eea
where c and s denotes the cosine and sine of the mixing angles, and
are given by

\bea{co}
 c = 1+O(V^{2}_\phi/V^{2}_\chi),\nonumber
 \eea \bea{si} s =
-\frac{\lambda_3-\lambda_4-\lambda_5}{2\lambda_1}(V_\phi/V_\chi)+O(V^{2}_\phi/V^{2}_\chi).
\eea
Since $V_{\phi} \sim 10^{-2}$ eV and $V_{\chi} \sim 250$ GeV,
this mixing is extremely small, and can be neglected.  Hence, we see
that $h$ behaves essentially like the SM Higgs  (except of course in
interactions with the neutrinos).

The interactions of the neutral Higgs fields with the $Z$ are given
by

\bea{Z_Higgs} \
L_{gauge}=\frac{\overline{g}}{2V}(cV_{\phi}+sV_{\chi})(\rho\partial^{\mu}h-h\partial^{\mu}\rho)Z_{\mu}\
+\frac{\overline{g}}{2V}(sV_{\phi}-cV_{\chi})(\rho\partial^{\mu}\sigma-\sigma\partial^{\mu}\rho)Z_{\mu}\nonumber\\
+\frac{\overline{g}^2}{4}(sV_{\phi}-cV_{\chi})hZ^{\mu}Z_{\mu}\
+\frac{\overline{g}^2}{4}(cV_{\phi}+sV_{\chi})\sigma Z^{\mu}Z_{\mu}\
+\frac{\overline{g}^2}{8}(h^{2}+\sigma^{2}+\rho^{2})Z^{\mu}Z_{\mu}
\eea
 where
$\overline{g}^{2}=g^{2}+g'^{2}$, and $V_{\chi}$ and $V_{\phi}$ are
the two VEV's.

\section{Phenomenological Implications}

We now consider the phenomenological implications of this model.
There are several interesting phenomenological implications which
can be tested in the upcoming neutrino experiments and high energy
colliders.  The light neutrinos in our model are Dirac particles. So
neutrinoless double beta decay is not allowed in our model.  This is
a very distinctive feature of our model for the neutrino masses
compared to the traditional see-saw mechanism.  In the see-saw
model, light neutrinos are Majorana particles, and thus neutrinoless
double beta decay is allowed.  The current limit on the double beta
decay is $m_{ee} \sim 0.3 ~eV$.  This limit is expected to go down
to about $m_{ee} \sim 0.01 ~eV$ in future experiments
\cite{betabeta}.  If no neutrinoless double beta decay is observed
to that limit, that will cast serious doubts on the see-saw model.
In our model, of course, it is not allowed at any level.

Next, we consider the implications of our model for high energy
colliders.  First we consider the production of the light scalar
$\sigma$ in $e^{+}e^{-}$ collisions.  The only possible decay modes
of this particle are a diphoton mode, $\sigma \rightarrow
\gamma\gamma$ which can occur at the one-loop level and, if it has
enough mass, a $\sigma \rightarrow \nu\overline{\nu}$ mode.  The one
loop decay to two photons takes place with quarks, $W$ bosons, or
charged Higgs bosons in the loop.  The largest contribution to this
decay mode is $\sim e^8 m_{\sigma}^5/{m_q}^4$. This gives the
lifetime of $\sigma$ to be $\sim 10^{20}$ years, which is much
larger than the age of the universe. Thus $\sigma$ essentially
behaves like a stable particle, and its  production at the colliders
will lead to missing energy in the event.  The couplings of $\sigma$
to quarks and charged leptons takes place only through mixing which
is highly suppressed (proportional to the ratio $V_\phi/V_\chi$).
Thus we need only consider its production via its interactions with
gauge bosons. The $ZZ\sigma$ coupling is also highly suppressed, so
that processes such as $e^{+}e^{-}\rightarrow Z^{*}\rightarrow
Z\sigma$ and $Z\rightarrow Z^{*}\sigma\rightarrow
f\overline{f}\sigma$ are negligible.  However, no such suppression
occurs for the $ZZ\sigma\sigma$ coupling.  Consider the $Z$ decay
process $Z\rightarrow Z^{*}\sigma\sigma\rightarrow
f\overline{f}\sigma\sigma$.  A direct calculation yields the width
(neglecting the $\sigma$ and fermion masses),

 \bea{width}
 \Gamma(Z\rightarrow f\overline{f}\sigma\sigma)
=
\frac{G^{3}_{F}m^{5}_{Z}(g^{2}_{V}+g^{2}_{A})}{2\sqrt{2}(2\pi)^{5}}\int^{m_{Z}/2}_{0}dE_1
\int^{m_{Z}/2}_{0}dE_2\nonumber\\ \times\int^{1}_{-1}d(\cos\theta)
\frac{E^{2}_{1}E^{2}_{2}(3-\cos\theta)}{(2E_{1}E_{2}-2E_{1}E_{2}\cos\theta-m^{2}_{Z})^{2}+m^{2}_{Z}\Gamma^{2}_{Z}},
\eea
where $g_{V} = T_{3}-2Q\sin^{2}\theta_{W}$ and $g_{A} = T_{3}$.
This gives

\bea{sum_width}\sum_{f}\Gamma (Z\rightarrow
f\overline{f}\sigma\sigma) \simeq 2.5\times10^{-7}~GeV. \eea
For the
$1.7\times10^{7}$ $Z$'s observed at resonance at LEP1 \cite{Z_res},
this gives an expectation of only about two such events.

Now we consider the production of the heavy Higgs particles in our
model. Since the charged Higgs $H^{\pm}$ and the pseudoscalar,
$\rho$ can be produced along with the light scalar $\sigma$, there
will be stricter mass bound on these particles than in a typical two
Higgs doublet model. Let us consider the pseudoscalar $\rho$, and
assume $m_\rho < m_Z$.  Then the Z can  decay via $Z\rightarrow
\sigma\rho$.
 Since $\rho$ couples negligibly to all SM fermions except the neutrinos,
  here we need only consider its decay to $\nu\overline{\nu}$
  (or $\sigma\sigma$ if we consider CP violation),  so this process
  contributes to the invisible decay width of the $Z$.  The width for this process is
\bea{width_rho} \Gamma =
\frac{G_{F}m^{3}_Z}{24\sqrt{2}\pi}\left(1-\frac{m^{2}_\rho}{m^{2}_Z}\right)^3
\eea This is less than the experimental uncertainty in the invisible
$Z$ width  for $m_\rho \gtrsim 78~GeV$. (The experimental value of
the invisible $Z$ width is $499.0\pm 1.5~MeV$ \cite{Z_width}.)

For $m_\rho > m_Z$, real pseudoscalar $\rho$ can be produced via
$e^{+}e^{-}\rightarrow Z^{*}\rightarrow\rho\sigma$.  The total cross
section for this process is

\bea{cro_sec} \sigma =
\frac{G^{2}_{F}m^{4}_{Z}(g^{2}_{V}+g^{2}_{A})s}{24\pi}\left(\frac{1}{s-m^{2}_{Z}}\right)^{2}\left(1-\frac{m^{2}_{\rho}}{s}\right)^3.
\eea For LEP2, $\sqrt{s}\simeq 200~GeV$, we find that less than one
event is expected in $\simeq3000~pb^{-1}$ \cite{higgs_search} of
data for $m_{\rho}\gtrsim 95~GeV$.  Note that the bound on the
$\rho$ mass we obtain is much less than the mass for which the Higgs
potential becomes strongly coupled ($\lambda_5 \le 2 \sqrt{\pi}$
which gives $m_{\rho} \le 470 $ GeV).

For $m_\rho > m_Z$, the $Z$ can still decay invisibly through
$Z\rightarrow \rho^{*}\sigma\rightarrow\nu\overline{\nu}\sigma$. The
width for this decay is

\bea{inv_width} \Gamma =
\frac{G_{F}m^{2}_{Z}y^{2}_{\nu_{l}}}{3\sqrt{2}(2\pi)^{3}}\int^{m_{Z}/2}_{0}dE
\frac{E^{3}(m_{Z}-2E)}{(m^{2}_{Z}-2m_{Z}E-m^{2}_{\rho})^{2}}. \eea
Summing over generations, this gives \bea{inv_1}
\Gamma(m_{\rho}=100~GeV) \simeq
(0.1~MeV)(\frac{1}{3}\sum_{l}y^{2}_{\nu_l})\nonumber \eea

\bea{inv_2}\Gamma(m_{\rho}=200~GeV) \simeq
(4\times10^{-3}~MeV)(\frac{1}{3}\sum_{l}y^{2}_{\nu_l}). \eea Even if
we take $\frac{1}{3}\sum y^{2}_{\nu_l}\sim1$, these values are well
within the experimental uncertainty in the invisible $Z$ width of
$1.5~MeV$.  Note that if we allow explicit CP violation in the Higgs
sector, the invisible decay
$Z\rightarrow\rho\sigma\rightarrow\sigma\sigma\sigma$ will also
occur.

Our model has very interesting implications for the discovery
signals of the Higgs boson at the high energy colliders, such as the
Tevatron and LHC.  Note that since $V_\phi$ is extremely small
compared to $V_\chi$, the neutral Higgs boson, $h$ is like the SM
Higgs boson so far its decays to fermions and to W and Z bosons are
concerned.  However, in our model, h has new decay modes, such as
$h\rightarrow\sigma\sigma$ which is invisible.  This could change
the Higgs signal at the colliders dramatically.  The width for this
invisible decay mode $h\rightarrow\sigma\sigma$ is given by

\bea{hig_wid_1}
 \Gamma(h\rightarrow\sigma\sigma) =
\frac{(\lambda_{3}+\lambda_{4}+\lambda_{5})^{2}V_{\chi}^2}{32\pi
m_{h}}.
\eea
Using
\bea{h_mass} m^{2}_{h} = 2\lambda_{1}V^{2}_{\chi}
+ O(V^{2}_\phi/V^{2}_\chi),
\eea
this can be written
\bea{hig_wid_2}\Gamma(h\rightarrow\sigma\sigma) =
\frac{(\lambda_{3}+\lambda_{4}+\lambda_{5})^{2}m_h}{64\pi
\lambda_{1}}.
\eea

\begin{figure} 
\centerline{
   \includegraphics[height=2.in]{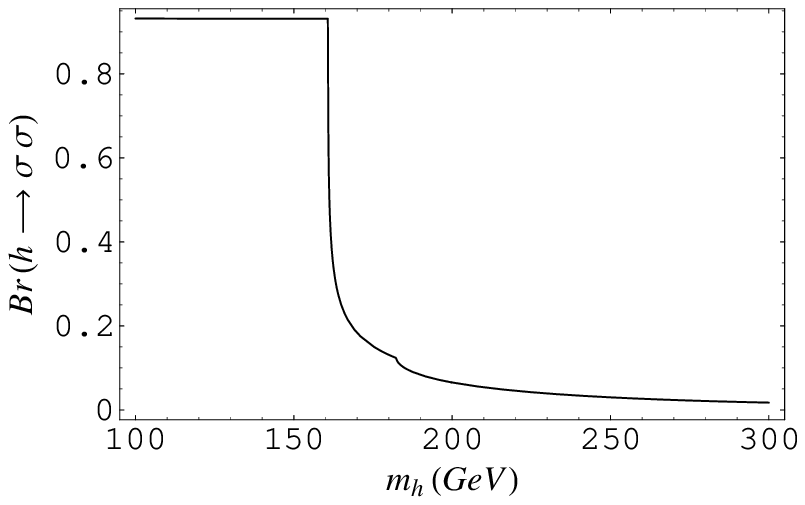}
   \includegraphics[height=2.in]{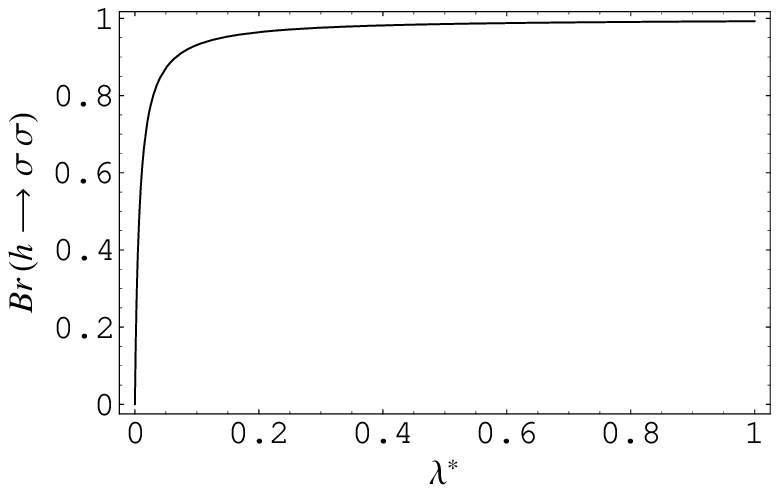}
   }
\caption{Left panel: Branching ratio for $h\rightarrow\sigma\sigma$
as a function of $m_h$ for the value of the parameter, $\lambda^* =
0.1$. Right panel: Branching ratio for $h\rightarrow\sigma\sigma$ as
a function of $\lambda^*$ for $m_h = 135$ GeV.}
\end{figure}

Depending on the parameters, it is possible for the dominant decay
mode of $h$ to be this invisible mode.  The branching ratios for the
Higgs decay to this invisible mode are shown in fig. 1 (left panel),
for the Higgs mass range from 100 to 300 GeV, for the choice of the
value of the parameter, $\lambda^*$ equal to $0.1$ where $\lambda^*$
is defined to be equal to
$\frac{(\lambda_{3}+\lambda_{4}+\lambda_{5})^{2}} {\lambda_{1}}$ .
Right panel in Fig. 1 shows how this branching ratio depends on this
parameter for a Higgs mass of 135 GeV. (The results for the
branching ratio is essentially the same for other values of the
Higgs mass between 120 and 160 GeV).  We see that for a wide range
of this parameter, for the Higgs mass up to about 160 GeV, the
invisible decay mode dominates, thus changing the Higgs search
strategy at the Tevatron run 2 and the LHC . The production rate of
the neutral scalar Higgs $h$ in our model are essentially the same
as in the SM. This implies that the Higgs mass bound from LEP is not
significantly altered .(The L3 collaboration set a bound of $m_h
\ge112.3$ GeV for an invisibly decaying Higgs with the SM production
rate \cite{l3}). However, because of the dominance of the invisible
decay mode, it will be very difficult to observe a signal at the LHC
in the usual production and decay channels such as $qqh
\rightarrow{qqWW}$, $qqh\rightarrow{qq\tau\tau}$,
$h\rightarrow{\gamma \gamma}$, $h\rightarrow{ZZ}\rightarrow {4l}$,
$t\overline{t}h$ (with $h\rightarrow b\overline{b}$) and
$h\rightarrow{WW}\rightarrow{l\nu l\nu l}$ \cite{atlas}. However, a
signal with such an invisible decay mode of the Higgs (as in our
model) can be easily observed at the LHC through the weak boson
fusion processes, $qq\rightarrow{qqW^{+} W^{-}}\rightarrow{qqH}$
 and $qq\rightarrow{qqZZ}\rightarrow{qqH}$ \cite{ez} if appropriate
 trigger could be designed for the ATLAS and CMS detector. For example,
 with only $10 fb^{-1}$ of data at the LHC, such a signal can be observed
 at the 95 percent CL with an invisible branching ratio of 31 percent or less for a Higgs
 mass of upto 400 GeV \cite{ez}. Thus our model can be easily tested at the LHC
 for a large region of the Higgs mass. Of course, establishing that this signal
 is from the Higgs boson production will be very difficult at the LHC. For the
 Higgs search at the Tevatron, the usual signal from the  $Wh$ production, and the
subsequent decays of h to $WW^*$ or $b\overline{b}$ will be absent.
The most promising mode in our model
 will be the production of ZH, with Z decaying to $l^{\pm}l^{\pm}$
 ($l = e,\mu$) and the Higgs decaying invisibly. There will be a
 peak in the missing energy distribution in the final state with a Z.
 We urge the Tevatron collaborations to look for such a signal.


Our model has several interesting astrophysical and cosmological
implications.  The scalar particle, $\sigma$ in our model is
extremely light, with mass $\sim 10^{-2} - 10^{-3} $ eV, but its
coupling to $\nu\overline{\nu}$ is unsuppressed.  This new
interaction will affect supernova explosion dynamics.  Also, since
this interaction $\nu\overline{\nu}\sigma$ can be pretty strong, it
can bound $\nu\overline{\nu}$ giving rise to $\nu\overline{\nu}$
atoms, thus giving rise to the possibility of new kind of star
formation. The spontaneous breaking of the discrete global symmetry,
$Z_2$ will lead to cosmological domain walls. It will be interesting
to see it can contribute to the vacuum energy in a significant way.
The scalar field, $\sigma$ in our model has also mass in the right
ballpark as the measured value of the cosmological constant, and it
is stable on a cosmological time scale.  It will be interesting to
see if it can be a viable candidate for the dark energy.
\section{Conclusions}

We have presented a simple extension of the Standard Model
supplemented by a discrete symmetry, $Z_2$.  We have also added
three right-handed neutrinos, one for each family of fermions, and
one additional Higgs doublet.  While the electroweak symmetry is
spontaneously broken at the usual 100 GeV scale, the discrete
symmetry, $Z_2$ remains unbroken to a scale of about $10^{-2}$ eV.
The spontaneous breaking of this $Z_2$ symmetry by the VEV of the
second Higgs doublet generates tiny masses for the neutrinos. The
neutrinos are Dirac particles in our model, so the neutrinoless
double beta decay is absent. This is a very distinctive feature of
our model for neutrino mass generation compared to the usual see-saw
model.  The neutral heavy Higgs in our model is very similar to the
SM Higgs in its couplings to the gauge bosons and fermions, but it
also couples to a very light scalar Higgs present in our model. This
light scalar Higgs, $\sigma$, is essentially stable, or decays to
$\nu\overline{\nu}$.  Thus the production of this $\sigma$ at the
high energy colliders leads to missing energy.  The SM-like Higgs,
for a mass up to about 160 GeV dominantly decays to the invisible
mode $h\rightarrow \sigma\sigma$.  Thus the Higgs signals at high
energy hadron colliders are dramatically altered in our model.  Our
model also has interesting implications for astrophysics and
cosmology.

\section*{Acknowledgments}

We thank K.S. Babu for useful discussions. We also thank D.
Rainwater for a very useful communication regarding the
observability of the Higgs signals in the presence of a invisible
decay mode. This work was supported in part by the US Department of
Energy, Grant Numbers DE-FG02-04ER41306 and DE-FG02-04ER46140.

\end{document}